\documentclass[12pt,cite,epsf,epsfig]{article}
\usepackage{epsfig}

\begin{document}

\author{S. Dev\thanks{Electronic address: dev5703@yahoo.com} ,
 Surender Verma\thanks{Electronic address: s\_7verma@yahoo.co.in} , Shivani
Gupta\thanks{Electronic address: shiroberts\_1980@yahoo.co.in}, R. R. Gautam\thanks{Electronic address: gautamrrg@gmail.com}}

\title{Neutrino Mass Matrices with a Texture Zero and a Vanishing Minor}
\date{\textit{Department of Physics, Himachal Pradesh University, Shimla 171005, INDIA.}\\
\smallskip}

\maketitle
\begin{abstract}
We study the implications of the simultaneous existence of a texture zero  and a vanishing
minor in the neutrino mass matrix. There are
thirty six possible texture structures of this type, twenty one of which reduce to
two texture zero cases which have, already, been extensively studied. Of the remaining fifteen textures only six are allowed by the current data. We
examine the phenomenological implications of the allowed texture structures for Majorana type CP-violating phases, 1-3 mixing angle and
Dirac type CP-violating phase. All these possible textures can be generated through the seesaw mechanism and realized in the framework of discrete abelian flavor symmetry. We present the symmetry realization of these texture structures.
\end {abstract}

\section{Introduction}
During last several years enormous progress has been made in
the determination of the neutrino masses and mixings and in studies
of the neutrino mass matrix. The main theoretical challenge is to
understand the dynamics behind the observed pattern of neutrino masses
and mixing. It is expected that detailed information of the
neutrino mass spectrum and lepton mixing may eventually shed light
on the origin of lepton masses, quark lepton symmetry and the
fermion mass problem. The main objectives of the neutrino physics include the determination of the absolute mass scale of neutrinos, their mass spectra/ mass hierarchy and also the subdominant structure of mixing: namely 1-3 mixing, deviation of 2-3 mixing from maximality and the CP violating phases. However, there exist a
large number of the possible structures of neutrino mass matrix.
Several proposals have been made in literature to restrict the
form of the neutrino mass matrix and to reduce the number of free
parameters which include presence of texture zeros
\cite{1,2,3,4,5, 6}, requirement of zero determinant \cite{7}, the
zero trace condition \cite{8}. In addition, the presence of vanishing minors \cite{9}, and the simultaneous
existence of a texture zero and an equality \cite{10} has been studied
in the literature. However, the current neutrino
oscillation data is consistent with only a limited number of texture
schemes. Detailed phenomenological analysis of the two texture
zeros \cite{1,2,3,4,5,6} has been done in the past. The seesaw
mechanism for understanding the scale of neutrino masses is
regarded as the prime candidate not only due to its simplicity but
also due to its theoretical appeal. In the framework of type I
seesaw mechanism \cite{11} the effective Majorana mass matrix
$M_{\nu}$ is given by
\begin{equation}
 M_{\nu}= - M_D M_R^{-1} M_D^T  \nonumber \\
\end{equation}
where $M_D$ is the Dirac neutrino mass matrix and $M_R$ is the
right handed Majorana mass matrix. It has been noted by many
authors \cite{12,13} that the zeros of the Dirac neutrino mass
matrix $M_D$ and the right handed
 Majorana mass matrix $M_R$ are the progenitors of zeros in the effective Majorana mass matrix $M_{\nu}$. Thus,
  the analysis of zeros in $M_D$ and $M_R$ is more basic than the study of zeros in $M_{\nu}$.
However, the zeros in $M_D$ and $M_R$ may not only show as zeros in effective neutrino mass matrix. Another
 interesting possibility is that these zeros show as a vanishing minor in the effective mass matrix $M_{\nu}$. Phenomenological analysis of the case where the zeros of $M_R$ show as a vanishing minor in $M_{\nu}$ for diagonal $M_D$ has been done recently \cite{9,13}. This, however, is not the most general case.
 In the present work we explore the more general possibility of simultaneous existence of a texture
  zero and a vanishing minor in $M_{\nu}$. Such texture structures are realized via seesaw mechanism when there
  are texture zeros in $M_D$ and $M_R$. Simultaneous existence of a texture zero and a vanishing minor restricts
the form of neutrino mass matrix and hence reduces the number of
free parameters to five. These texture structures
 can be generated through the seesaw mechanism and realized by a discrete abelian flavor symmetry \cite{14}. We present
 the detailed analysis for such texture structures and examine their phenomenological
implications. It is found that there are thirty six such
 texture structures twenty one of which reduce to two texture zero form which have been extensively studied in the literature. We analyse the remaining fifteen texture
structures and find that only six structures are consistent with
the available data. We study the phenomenological
implications for these texture structures and present their symmetry realization.

\section{Neutrino mass matrix}
We reconstruct the neutrino mass matrix in the flavor basis (where
the charged lepton mass matrix is diagonal)
 assuming also that neutrinos are Majorana particles.
In this basis a complex symmetric neutrino mass matrix can be
diagonalized by a unitary matrix $V$ as
\begin{equation}
M_{\nu}=VM_{\nu}^{diag}V^{T}.
\end{equation}
where
\begin{center}
$M_{\nu}^{diag}$ = $\left(
\begin{array}{ccc}
m_{1} & 0 & 0 \\  0& m_{2} & 0 \\ 0& 0 & m_{3}
\end{array}
\right)$
\end{center}
The  matrix $M_{\nu}$ can be parameterized in terms of three
neutrino mass eigenvalues ($m_1, m_2, m_3$), three neutrino mixing
angles ($\theta _{12}$, $\theta _{23}$, $\theta _{13}$) (solar,
atmospheric and the reactor neutrino mixing angles respectively) and the Dirac
type CP- violating phase $\delta$. The two additional phases
$\alpha$ and $\beta$ appear if neutrinos are Majorana particles.
Here the matrix
\begin{equation}
V = UP
\end{equation}
where
\begin{equation}
U= \left(
\begin{array}{ccc}
c_{12}c_{13} & s_{12}c_{13} & s_{13}e^{-i\delta} \\
-s_{12}c_{23}-c_{12}s_{23}s_{13}e^{i\delta} &
c_{12}c_{23}-s_{12}s_{23}s_{13}e^{i\delta} & s_{23}c_{13} \\
s_{12}s_{23}-c_{12}c_{23}s_{13}e^{i\delta} &
-c_{12}s_{23}-s_{12}c_{23}s_{13}e^{i\delta} & c_{23}c_{13}
\end{array}
\right)
\end{equation} with $s_{ij}=\sin\theta_{ij}$ and $c_{ij}=\cos\theta_{ij}$ and
\begin{center}
$P = \left(
\begin{array}{ccc}
1 & 0 & 0 \\ 0 & e^{i\alpha} & 0 \\ 0 & 0 & e^{i(\beta+\delta)}
\end{array}
\right)$
\end{center} is the diagonal phase matrix with
the two Majorana type CP- violating phases $\alpha$, $\beta$ and Dirac type CP- violating phase $\delta$.
 The matrix $V$ is called the neutrino mixing matrix or the Pontecorvo-Maki-Nakagawa-Sakata (PMNS) matrix  \cite{15}.
  Using Eq. (3) and Eq. (4), the neutrino mass matrix can be written as
\begin{equation}
M_{\nu}=U P M_{\nu}^{diag}P^{T}U^{T}.
\end{equation}
The CP violation in neutrino oscillation experiments can be described
through a rephasing invariant quantity, $J_{CP}$ \cite{16} with
$J_{CP}=Im(U_{e1}U_{\mu2}U_{e2}^*U_{\mu1}^*)$. In our
parameterization, $J_{CP}$ is given by
\begin{equation}
J_{CP} = s_{12}s_{23}s_{13}c_{12}c_{23}c_{13}^2 \sin \delta.
\end{equation}
The observation of  neutrinoless double beta decay would signal lepton number violation and imply
 Majorana nature of neutrinos. The effective Majorana mass of the electron neutrino $M_{ee}$ which
 determines the rate of neutrinoless double beta decay is given by
\begin{equation}
M_{ee}= |m_1c_{12}^2c_{13}^2+ m_2s_{12}^2c_{13}^2 e^{2i\alpha}+ m_3s_{13}^2e^{2i\beta}|.
\end{equation}

The experimental constraints on neutrino parameters at 1, 2 and
3$\sigma$ \cite{17} are given below:
\begin{eqnarray}
\Delta m_{12}^{2}
&=&7.67_{(-0.19,-0.36,-0.53)}^{(+0.16,+0.34,+0.52)}\times
10^{-5}eV^{2}, \nonumber \\ \Delta m_{23}^{2} &=&\pm
2.39_{(-0.8,-0.20,-0.33)}^{(+0.11,+0.27,+0.47)}\times 10^{-3}eV^{2},  \nonumber \\
\theta_{12}& =&33.96_{(-1.12,-2.13,-3.10)}^{o(+1.16,+2.43,+3.80)},
\nonumber \\ \theta_{23}
&=&43.05_{(-3.35,-5.82,-7.93)}^{o(+4.18,+7.83,+10.32)}, \nonumber \\
\theta_{13} &<& 12.38^{o}(3\sigma).
\end{eqnarray}
The upper bound on $\theta_{13}$ is given by the CHOOZ experiment.

\section{Neutrino mass matrices with one texture zero and one vanishing minor}
The simultaneous existence of a texture zero and a vanishing minor in the neutrino mass matrix gives,
\begin{center}
 \begin{eqnarray}
M_{\nu (xy)}=0, \\ \nonumber M_{\nu (pq)} M_{\nu (rs)}- M_{\nu
(tu)} M_{\nu (vw)}=0.
\end{eqnarray}
 \end{center}
These two conditions  yield two complex equations viz.
\begin{equation}
m_{1}X+ m_{2}Y e^{2i\alpha
}+m_{3}Z e^{2i(\beta +\delta )}=0,
\end{equation}
where $X = U_{x1}U_{y1}, Y = U_{x2}U_{y2}, Z = U_{x3}U_{y3}$ and
\begin{equation}
\sum_{l,k=1}^{3}(U_{pl}U_{ql}U_{rk}U_{sk}-U_{tl}U_{ul}U_{vk}U_{wk})m_{l}m_{k}=0
\end{equation}
We denote the minor corresponding to the $(ij)^{th}$ element by
$C_{ij}$. The equation of vanishing minor becomes
\begin{equation}
m_1 m_2 A_3e^{2i\alpha} + m_2 m_3 A_1e^{2i(\alpha+\beta +\delta )}+m_3 m_1A_2e^{2i(\beta +\delta)}=0
\end{equation}
where
\begin{equation}
A_h=(U_{pl}U_{ql}U_{rk}U_{sk}-U_{tl}U_{ul}U_{vk}U_{wk})+(l\leftrightarrow k)
\end{equation}
with $(h,l,k)$ as the cyclic permutation of (1,2,3). These two complex eqns. (10) and (12) involve nine physical parameters $m_{1}$, $m_{2}$, $m_{3}$,
$\theta _{12}$, $\theta _{23}$, $\theta _{13}$ and three
CP-violating phases $\alpha $, $\beta $ and $\delta $. The masses
$m_{2}$ and $m_{3}$ can be calculated from the mass-squared
differences $\Delta m_{12}^{2}$ and $\Delta m_{23}^{2}$ using the
relations
\begin{equation}
m_{2}=\sqrt{m_{1}^{2}+\Delta m_{12}^{2}},
\end{equation}
and
\begin{equation}
m_{3}=\sqrt{m_{2}^{2}+\Delta m_{23}^{2}}.
\end{equation}
 Using the experimental
inputs of the two mass squared
 differences and the two mixing angles we can constrain the other parameters.
  Thus, in the above two complex equations we are left with five unknown parameters
   $m_1$, $\alpha$, $\beta$, $\delta$ and $\theta_{13}$ which are, obviously, correlated. Simultaneously solving Eqs. (10) and (12) for the two mass ratios , we obtain
\begin{equation}
\frac{m_1}{m_3}e^{-2i\beta }=-\frac{%
\left( X A_1-Y A_2+Z A_3\pm
\sqrt{X^2A_1^2+(YA_2-ZA_3)^2-2XA_1(YA_2+ZA_3)}\right)}{2 X
A_3}e^{2i\delta },
\end{equation}
and
\begin{equation}
\frac{m_1}{m_2}e^{-2i\alpha }=\frac{%
\left( -X A_1- YA_2+ ZA_3\pm
\sqrt{X^2A_1^2+(YA_2-ZA_3)^2-2XA_1(YA_2+ZA_3)}\right)}{2 X A_2}.
\end{equation}
The magnitude of the two mass ratios is given as
\begin{equation}
\rho=\left|\frac{m_1}{m_3}e^{-2i\beta }\right|,
\end{equation}
\begin{equation}
\sigma=\left|\frac{m_1}{m_2}e^{-2i\alpha }\right|.
\end{equation}
 while the CP- violating Majorana phases $\alpha$ and $\beta$ are given by
\begin{equation}
\alpha =-\frac{1}{2}arg\left(\frac{%
\left( -X A_1- YA_2+ ZA_3\pm
\sqrt{X^2A_1^2+(YA_2-ZA_3)^2-2XA_1(YA_2+ZA_3)}\right)}{2 X
A_2}\right),
\end{equation}
\begin{equation}
\beta =-\frac{1}{2}arg\left(-\frac{%
\left( X A_1-YA_2+ZA_3\pm
\sqrt{X^2A_1^2+(YA_2-ZA_3)^2-2XA_1(YA_2+ZA_3)}\right)}{2 X
A_3}e^{2i\delta }\right).
\end{equation}
Since, $\Delta m_{12}^{2}$ and $\Delta m_{23}^{2}$ are known experimentally, the values of mass ratios $(\rho,\sigma)$ from Eq. (18) and (19) can be used to calculate $m_1$.
This can be done by inverting Eqs. (14) and (15) to obtain the two values of $m_1$, viz.
\begin{equation}
m_{1}=\sigma \sqrt{\frac{ \Delta
m_{12}^{2}}{1-\sigma ^{2}}},
\end{equation}
and
\begin{equation}
m_{1}=\rho \sqrt{\frac{\Delta m_{12}^{2}+
\Delta m_{23}^{2}}{ 1-\rho^{2}}}.
\end{equation}
We vary the oscillation parameters within their known experimental ranges. However, the Dirac type CP- violating phase $\delta$ is varied within its full range and $\theta_{13}$ is varied in its 3$\sigma$ range given by the CHOOZ bound. The two values of $m_1$ obtained from the mass ratios $\rho$ and $\sigma$, respectively must be equal to within the errors of the oscillation parameters for the simultaneous existence of a texture zero and a vanishing minor. 

 There are in total thirty six possible structures of neutrino mass matrix [Table 1.]
   with a single texture zero and a vanishing minor. As can be seen from Table 1., twenty one structures corrosponds to two texture zero cases which have, already, been studied extensively.
   We examine the phenomenological viability of all the remaining texture structures and also present detailed phenomenological implications for the viable structures.

\begin{table}[h]
\begin{center}
\begin{small}
\begin{tabular}{|c|c|c|c|c|c|c|c|}
\hline  &  & A. & B. &C. & D. & E.& F. \\
\hline 1. & $\left(
\begin{array}{ccc}
0 & b & c \\  b & d & e \\ c& e & f
\end{array}
\right)$ & $df - e^{2}=0$ & $bf-ec=0$ & $be-cd=0$ & Two Zero & Two Zero& Two Zero \\
\hline 2. &$\left(
\begin{array}{ccc}
a & 0 & c \\  0& d & e \\ c& e & f
\end{array}
\right)$   & $df-e^2=0$ & Two Zero &Two Zero  & $af-c^2=0$ & Two Zero&Two Zero \\
\hline 3. & $\left(
\begin{array}{ccc}
a & b & 0 \\ b & d & e \\0 &e  & f
\end{array}
\right)$ &$df-e^2=0$  & Two Zero & Two Zero & Two Zero &Two Zero  & $ad-b^2=0$\\
\hline 4. & $\left(
\begin{array}{ccc}
a & b & c \\ b & 0 & e \\c & e & f
\end{array}
\right)$ & Two Zero &$bf-ec=0$  & Two Zero &$af-c^2=0 $ & $ae-bc=0$&Two Zero \\
\hline 5. &$\left(
\begin{array}{ccc}
a & b & c \\  b & d & 0 \\ c & 0 & f
\end{array}
\right)$  & Two Zero &Two Zero  & Two Zero & $af-c^2=0$ & Two Zero &$ad-b^2=0$\\
\hline 6. & $\left(
\begin{array}{ccc}
a & b & c \\ b & d & e \\ c& e & 0
\end{array}
\right)$ & Two Zero &Two Zero  & $be-dc=0$ & Two Zero &$ae-bc=0$  &$ad-b^2=0$\\
\hline
\end{tabular}
\end{small}
\caption{Thirty Six allowed texture structures of $M_{\nu}$ with a texture Zero and one vanishing minor.}
\end{center}
\end{table}
\newpage
\section{Results and Discussion}
As pointed out earlier, the two values of $m_1$ obtained from Eq. (22) and Eq. (23) must be equal, of course, to within the errors of the oscillation parameters, for the simultaneous existence of a texture zero and a vanishing minor. Correspondingly, we obtain two regions of solutions. The viability of the simultaneous existence of a texture zero and a vanishing minor in the neutrino mass matrix is studied for both these regions. The known oscillation parameters are varied within their experimental ranges while the Dirac type CP- violating phase $\delta$ is varied within its full possible range. The 1-3 mixing angle is varied over the range limited by the 3$\sigma$ CHOOZ bound. It is found that out of the fifteen possible structures of the neutrino mass matrix with a texture zero and a vanishing minor, only six are allowed at 99\% C.L. while the remaining nine are phenomenologically disallowed. We examine the phenomenological implications of all the six viable structures. We present the lower bound on the effective Majorana mass, $M_{ee}$ for the viable cases. This important parameter determines the rate of neutrinoless double beta decay and will help decide the nature of neutrinos. The analysis of $M_{ee}$ will be significant as many neutrinoless
double beta decay experiments will constrain this
parameter. A most stringent constraint on the value of $M_{ee}$
was obtained in the $^{76}Ge$
 Heidelberg-Moscow experiment \cite{18} $|M_{ee}|< 0.35$eV.
 There are large number of projects such as SuperNEMO\cite{19},
 CUORE\cite{20}, CUORICINO\cite{20} and  GERDA\cite{21} which
 aim to achieve a sensitivity below 0.01eV to $M_{ee}$.
 Forthcoming experiment SuperNEMO, in particular, will explore
 $M_{ee}$ $<$ 0.05eV\cite{22}.
In addition we also predict the bounds on 1-3 mixing angle, and
deviation of 2-3 mixing angle from maximality for some viable
cases. The precise measurements of the mixing angles and in
particular, searches for the deviations of 1-3 mixing from zero
and 2-3 mixing from maximality, is crucial for understanding the
underlying physics. The proposed experiments such as Double CHOOZ
plan to explore $\sin^22\theta_{13}$ down to 0.06 in phase I (0.03
in phase II) \cite{23}. Daya Bay has a higher sensitivity and plans
to observe $\sin^22\theta_{13}$ down to 0.01 \cite{24}.
Next, we present the detailed numerical analysis for the six viable texture structures of neutrino mass matrix with a
texture zero and a vanishing minor. In addition we study class 2A
and 3A analytically as these two classes give strongly hierarchical mass spectrum (IH). We also present the correlation plots
between different parameters at 3$\sigma$ C.L.,
  and obtain interesting constraints on some of the parameters which are testable in the near future.
   We generate about $10^6$ random points in our numerical analysis, and make direct use
    of the two mass squared differences (Eqs. 22, 23), thus, making our analysis more reliable.
\subsection{Class 2A.}
This texture structure has zero (1,2) element and zero minor
corresponding to (1,1) entry ($C_{11}=0$). This class
has a clear inverted hierarchy. Here,
 $A_1 = c_{12}^2c_{13}^2,
 A_2=  c_{13}^2s_{12}^2,
 A_3= s_{13}^2 e^{2i\delta}.$
 We obtain the following analytical approximations for the mass ratios in the leading order of $s_{13}$ as
 \begin{equation}
\rho= \left|\frac{m_1}{m_3}\right|\approx  \frac{1}{s_{13}^2} + O (\frac{1}{s_{13}}),
\end{equation}
\begin{equation}
\sigma= \left|\frac{m_1}{m_2}\right|\approx  1-\frac{cos\delta s_{13}
s_{23}}{c_{12} c_{23} s_{12}}+ O (s_{13}^2).
\end{equation}
It can be seen from Eq. (24) that $\rho$ is always greater than 1,
which gives inverted mass hierarchy of neutrino masses.
Since, the mass ratio $\sigma$ is always smaller
than one we find from Eq. (25) that cos$\delta$ is always positive i.e. $\delta$ lies in the first and fourth quadrant.
 Fig. 1. gives the correlation plots for this texture. There exists a lower bound on
 effective Majorana neutrino mass $M_{ee} >$ 0.042 eV and $\theta_{13}> 0.35^o$ .We get highly constrained parameter space for
  the two Majorana type CP - violating phases $\alpha$ and $\beta$ (Fig. 1(b)). Larger values of
$\theta_{13}$ are allowed for $\delta $ near $90^o, 270^o$ as seen
from Fig. 1(d).  The Jarlskog rephasing invariant $J_{CP}$ is
 within the range (-0.45)-(0.45) (Fig. 1(c)).

\subsection{Class 3A.}

This texture structure has zero 13 element and zero minor
corresponding to 11 entry ($C_{11}=0$). Class 3A  also gives clear
inverted mass hierarchy of neutrino masses. This texture has same
values of $A_1$, $A_2$ and $A_3$ as for Class 2A since both have
the same vanishing minor. The analytical approximations for the mass
ratios in the leading order of $s_{13}$ are given as
 \begin{equation}
\rho= \left|\frac{m_1}{m_3}\right|\approx  \frac{1}{s_{13}^2} + O (\frac{1}{s_{13}}),
\end{equation}
\begin{equation}
\sigma= \left|\frac{m_1}{m_2}\right|\approx 1 + \frac{cos\delta s_{13}
c_{23}}{c_{12} s_{23} s_{12}}+ O (s_{13}^2).
\end{equation}
It can be seen from Eq. (26) that $\rho$
is always greater than 1, which gives inverted hierarchy of
neutrino masses. Since, the mass ratio
$\sigma$ is always smaller than one we find
from Eq. (27) that cos$\delta$ is always negative i.e. $\delta$
lies in the second and third quadrant. Fig. 2. gives some
interesting correlation plots for this texture structure. A lower
bound on $M_{ee}>$ 0.044 eV and 1-3 mixing angle
($\theta_{13}>0.3^o$) is obtained for this class. Larger values of
$\theta_{13}$ are allowed near $90^o, 270^o$. The rephasing
invariant, $J_{CP}$ lies in the range (-0.045)-(0.045). 

It is found that the limit of a vanishing mass eigenvalue i.e. $m_3$=0 can be reached for both class 2A and 3A. As can be seen from Fig. 2(b) that very small range ($0.3^o\le$$\theta_{13}$$\le1.1^o$) is allowed for the limit of vanishing mass eigenvalue $m_3$=0.

 It is interesting to note that class 3A is transformed to class 2A and vice-versa by the transformation $\delta\rightarrow\delta+\pi$ , $\theta_{23}\rightarrow(\frac{\pi}{2}-\theta_{23})$. Therefore, the predictions for neutrino mass matrices for these two classes will be identical for all neutrino parameters except $\theta_{23}$ and/ or $\delta$.

\subsection{Remaining Viable Classes.} The remaining viable texture
structures which have phenomenological implications are 2D, 3F,
4B, 6C. The corresponding values of $A_1$, $A_2$ and $A_3$ for
these texture structures are given in Table 2. All these textures
give normal, inverted and quasidegenerate mass spectra. For class
2D(NH, QD), we get unconstrained parameter space i.e. there are no
strong predictions. However, for 2D(IH), the Dirac type CP-
violating phase $\delta$ is constrained to the range $90^o -
270^o$, while the Majorana type CP- violating phase $\alpha$ takes
the value $0^o, 180^o$. A lower bound on the Effective Majorana
mass $M_{ee}> 0.05$ eV is obtained. The atmospheric mixing angle,
$\theta_{23}$ lies below maximality i.e. $\theta_{23}< 45^o$
(Fig.3(a)). Class 3F(IH) has similar predictions and bounds for
all parameters as in class 2D(IH) except $\theta_{23}$ which is
above maximal. For class 4B and 6C, unconstrained parameter space
is obtained for inverted and quasidegenerate mass hierarchy.
However, 4B(NH) and 6C(NH) give some strong predictions for
parameters under investigation. A lower bound $M_{ee}> 0.01$ eV is
obtained for both these cases. The atmospheric mixing angle
$\theta_{23}$ is above maximal for 6C(NH) and below maximal for
4B(NH).  There are some projects like Tokai- to- Kamioka-
 Korea (T2KK) which plans to resolve the octant degeneracy of
 $\theta_{23}$ (i.e. $\theta_{23}<45^{o}  $ or $\theta_{23}>45^{o}$) \cite{25}.
\begin{table}[h]
\begin{center}
\begin{tabular}{|c|c|c|c|}
\hline  Texture & $A_1$& $A_2$ & $A_3$  \\
\hline 2D.
&$e^{-2i\delta}(e^{i\delta}s_{12}c_{23}+s_{13}s_{23}c_{12})^2$
 &$e^{-2i\delta}(e^{i\delta}c_{12}c_{23}-s_{13}s_{23}s_{12})^2$  &$c_{13}^2s_{23}^2$    \\
\hline 3F. & $e^{-2i\delta}(-e^{i\delta}s_{12}s_{23}+s_{13}c_{23}c_{12})^2$ & $e^{-2i\delta}(e^{i\delta}c_{12}s_{23}+s_{13}c_{23}s_{12})^2$  & $c_{13}^2c_{23}^2$  \\
\hline 4B. & $c_{12}c_{13}(e^{-i\delta}c_{12}s_{23}s_{13}+ c_{23}s_{12})$ &  $s_{12}c_{13}(e^{-i\delta}s_{12}s_{23}s_{13}-c_{23}c_{12})$ &$-e^{i\delta}(c_{13}s_{23}s_{13})$   \\
\hline 6C. & $c_{12}c_{13}(-e^{-i\delta}c_{12}c_{23}s_{13}+s_{23}s_{12})$ &$s_{12}c_{13}(-e^{-i\delta}s_{12}c_{23}s_{13}-s_{23}c_{12})$ & $e^{i\delta} s_{13}c_{23}c_{13}$  \\
\hline
\end{tabular}
\caption{$A_1$, $A_2$ and $A_3$ for the remaining viable textures.}
\end{center}
\end{table}
\section{Symmetry Realization}
All the phenomenologically viable textures with a texture zero and a
zero minor in $M_\nu$ discussed in this work can be realized in a
simple way in models based on seesaw mechanism with a
abelian flavor symmetry \cite{14}. For constructing the required leptonic
mass matrices, we consider three left-handed Standard Model (SM)
lepton doublets $D_{La}$(a=1,2,3) and three right handed charged
lepton singlets $l_{Ra}$. To this we add three right handed
neutrino singlets $\nu_{Ra}$ in order to enable the see-saw mechanism for
suppressing the neutrino masses. For each non zero entry in $M_l$
or $M_D$ we need one Higgs doublet. Similarly, for each non-zero
matrix element of $M_R$ we need a scalar singlet $\chi_{ab}$.
Here, we present the symmetry realization for the
texture structure 2D which can be generated through seesaw mechanism.
One of the possible texture structure of  $M_D$ and $M_R$ which
reproduce class 2D is
\begin{equation}
M_D=\left(
\begin{array}{ccc}
0 & 0 & a \\ b & c & 0 \\ 0 & d & e
\end{array}
\right),M_R=
\left(
\begin{array}{ccc}
0& A & 0 \\ A & B & 0 \\ 0 & 0 & C
\end{array}
\right),
\end{equation}
\nonumber
The resulting $M_{\nu}$ generated through seesaw mechanism, takes the form
\begin{equation}
M_{\nu}=\left(
\begin{array}{ccc}
\matrix{ \frac{a^2}{C} & 0 & \frac{a\,e}{C} \cr 0 & \frac{b\,
\left( -\left( b\,B \right)  + 2\,A\,c \right) }{A^2} &
\frac{b\,d}{A} \cr \frac{a\,e} {C} & \frac{b\,d}{A} &
\frac{e^2}{C} \cr  }
\end{array}
\right),
\end{equation}
This $M_{\nu}$ has a zero  (1,2) element and a zero minor
corresponding to the (2,2) element $(C_{22}=0)$ as required. The symmetry
realization for this texture structure could be done through a generic
choice of the abelian symmetry group \textbf{$Z_{12}\times Z_2$}
discussed in \cite{14} which however may not be the most
economic way. Under \textbf{$Z_{12}$} the leptonic fields
transform as
\begin{center}
\begin{eqnarray}
\overline{l}_{R1}\rightarrow \omega \overline{l}_{R1},&  \overline{\nu}_{R1}\rightarrow \omega \overline{\nu}_{R1},&  D_{L1}\rightarrow  \omega D_{L1},  \nonumber \\  \overline{l}_{R2}\rightarrow \omega^2 \overline{l}_{R2}, & \overline{\nu}_{R2}\rightarrow \omega^2 \overline{\nu}_{R2}, & D_{L2}\rightarrow \omega^3 D_{L2}, \\ \overline{l}_{R3}\rightarrow \omega^5\overline{l}_{R3}, & \overline{\nu}_{R3}\rightarrow \omega^5\overline{\nu}_{R3},&  D_{L3}\rightarrow \omega^8D_{L3}. \nonumber
\end{eqnarray}
\end{center}
where $\omega$ = exp($i \pi/6$).
The bilinears $\overline{l}_{Ra}D_{Lb}$  and  $\overline{\nu}_{Ra}D_{Lb}$, relevant for $(M_l)_{ab}$ and $(M_D)_{ab}$ transform as
\begin{center}
$\left(
\begin{array}{ccc}
\omega^2& \omega^4 &\omega^9 \\
\omega^3 & \omega^5 &\omega^{10} \\
\omega^6& \omega^8& \omega
\end{array}
\right)$
\end{center}
while the bilinears $\overline{\nu}_{Ra} C
\overline{\nu}_{Rb}^T$, relevant for $(M_R)_{ab}$, transform as
\begin{center}
$\left(
\begin{array}{ccc}
\omega^2 & \omega^3 & \omega^6 \\ \omega^3 & \omega^4 &\omega^7 \\ \omega^6 & \omega^7& \omega^{10}
\end{array}
\right)$
\end{center}
To obtain diagonal charged lepton mass matrix, only three Higgs doublets are needed viz. $\phi_{11}$, $\phi_{22}$
and $\phi_{33}$. Under $Z_{12}$ these scalar doublets get respectively multiplied by $\omega^{10}$, $\omega^{7}$
and $\omega^{11}$ so that the charged lepton mass term remains invariant. The non-diagonal entries of $M_l$ remain
zero in the absence of any further Higgs doublets. Similarly non-zero entries of $M_D$ and $M_R$ in Eq.(28) can be
 obtained by introducing scalar Higgs doublets $\tilde{\phi_{13}}$, $\tilde{\phi_{21}}$, $\tilde{\phi_{22}}$, $\tilde{\phi_{32}}$ and
  $\tilde{\phi_{33}}$ being multiplied by $\omega^{3}$, $\omega^9$, $\omega^7$, $\omega^{4}$ and $\omega^{11}$ respectively
  under $Z_{12}$ for $M_D$ and by introducing scalar  singlet fields namely $\chi_{12}$, $\chi_{22}$ and $\chi_{33}$
  which get multiplied by $\omega^{9}$, $\omega^{8}$ and $\omega^{2}$ under $Z_{12}$ for $M_R$. It is important to
   note that the scalar Higgs doublets acquire vacuum expectation values (vev) at the electroweak scale, while scalar
   singlets acquire vevs at the seesaw scale. Under $Z_2$ the $\tilde{\phi_{ab}}$ and the neutrino singlets $\nu_{Ra}$
   change sign, while all other multiplets remain invariant. The symmetry realization for different $M_D$ and $M_R$
 giving $M_{\nu}$ corresponding to our viable textures can be similarly performed.

\section{Conclusions}
We presented a comprehensive phenomenological analysis for the Majorana neutrino mass matrices with a texture zero and a vanishing minor. All these texture structures can be generated through seesaw mechanism when there are texture zeros in $M_D$ and $M_R$ and realized in the framework of discrete abelian flavor symmetry. It is found that out of a total of thirty six texture structures, twenty one reduce to two zero texture structures which have been extensively studied in the past. The viability of the simultaneous
 existence of a texture zero and a vanishing minor in the neutrino mass matrix is studied for the two regions of solutions.
  Nine out of remaining textures are disallowed by the current data and we presented the numerical analysis for the remaining six texture structures. Analytical framework for the two classes with strongly hierarchical mass spectrum is, also, given. Predictions for 1-3 mixing angle and the Dirac type CP- violating phase are given for the allowed texture structures. These parameters are expected to be measured in the forthcoming neutrino oscillation experiments. We, also, obtained the lower bound on the effective Majorana mass for different classes. In the end, we presented the symmetry realization of these texture structures which are generated via the seesaw mechanism. However, the evolution of the Yukawa coupling matrices of the mass operators in models with multi Higgs doublets is an important issue since the evolution of the coupling matrices from the seesaw scale down to the electroweak scale may alter the predictions of the models under consideration. However, these corrections, in general, are expected to be neglegibly small except perhaps for the case of a degenerate neutrino mass spectrum which needs to be investigated carefully.

\textbf{\textit{\Large{Acknowledgements}}}

The research work of S. D. is supported by the University Grants
Commission, Government of India \textit{vide} Grant No. 34-32/2008
(SR).  S. G., R. R. G. and S. V. acknowledges the financial support provided
by Council for Scientific and Industrial Research (CSIR) and
University Grants Commission (UGC), Government of India,
respectively.

\newpage
\begin{figure}
\begin{center}
\epsfig{file=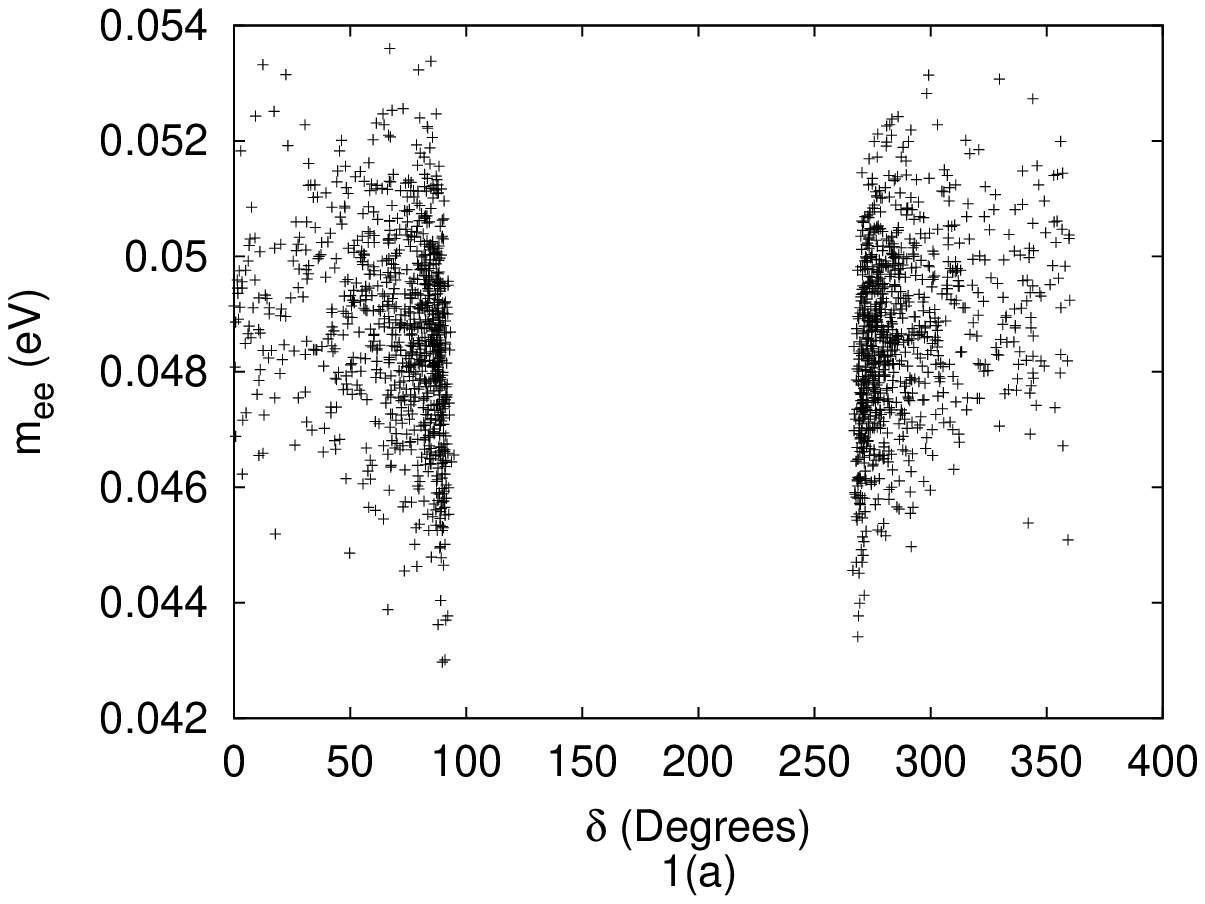,height=5.0cm,width=5.0cm}
\epsfig{file=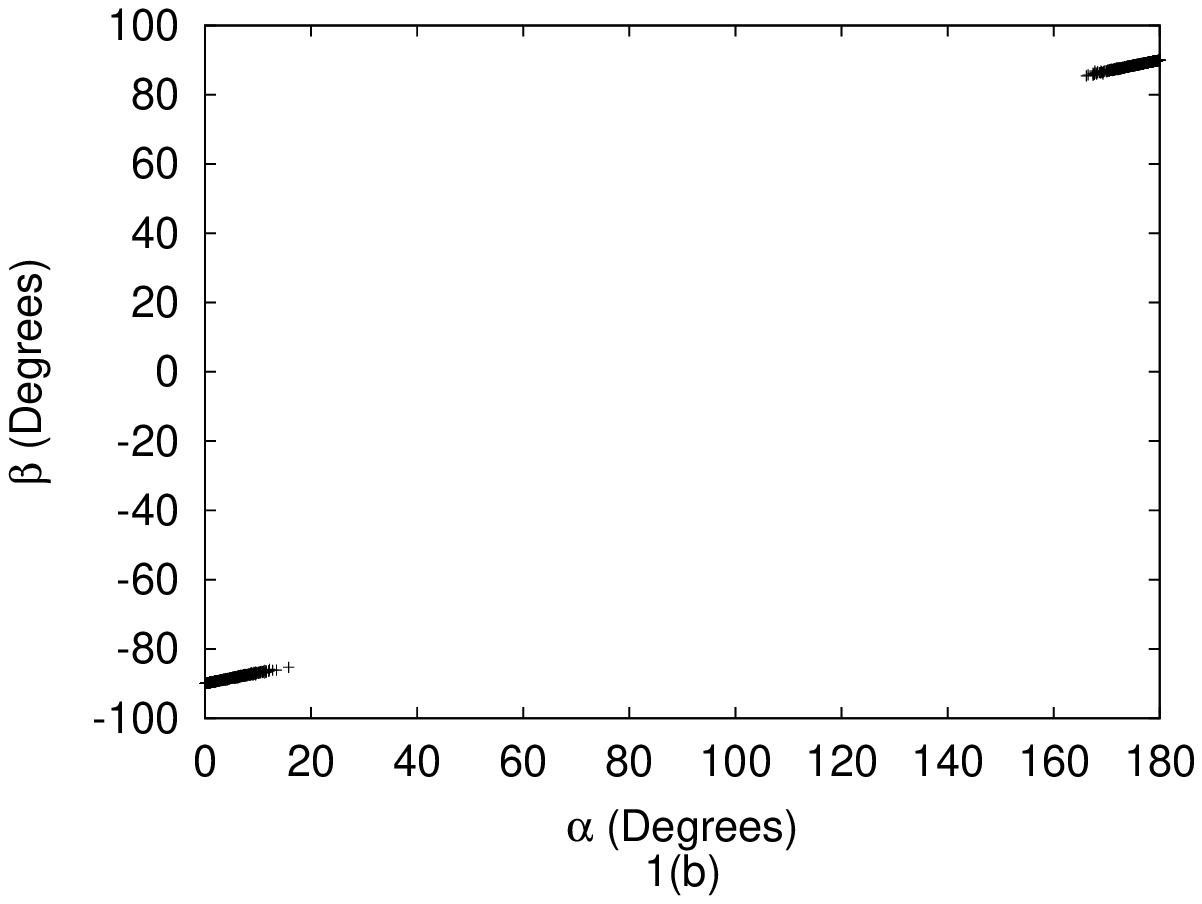,height=5.0cm,width=5.0cm}
\epsfig{file=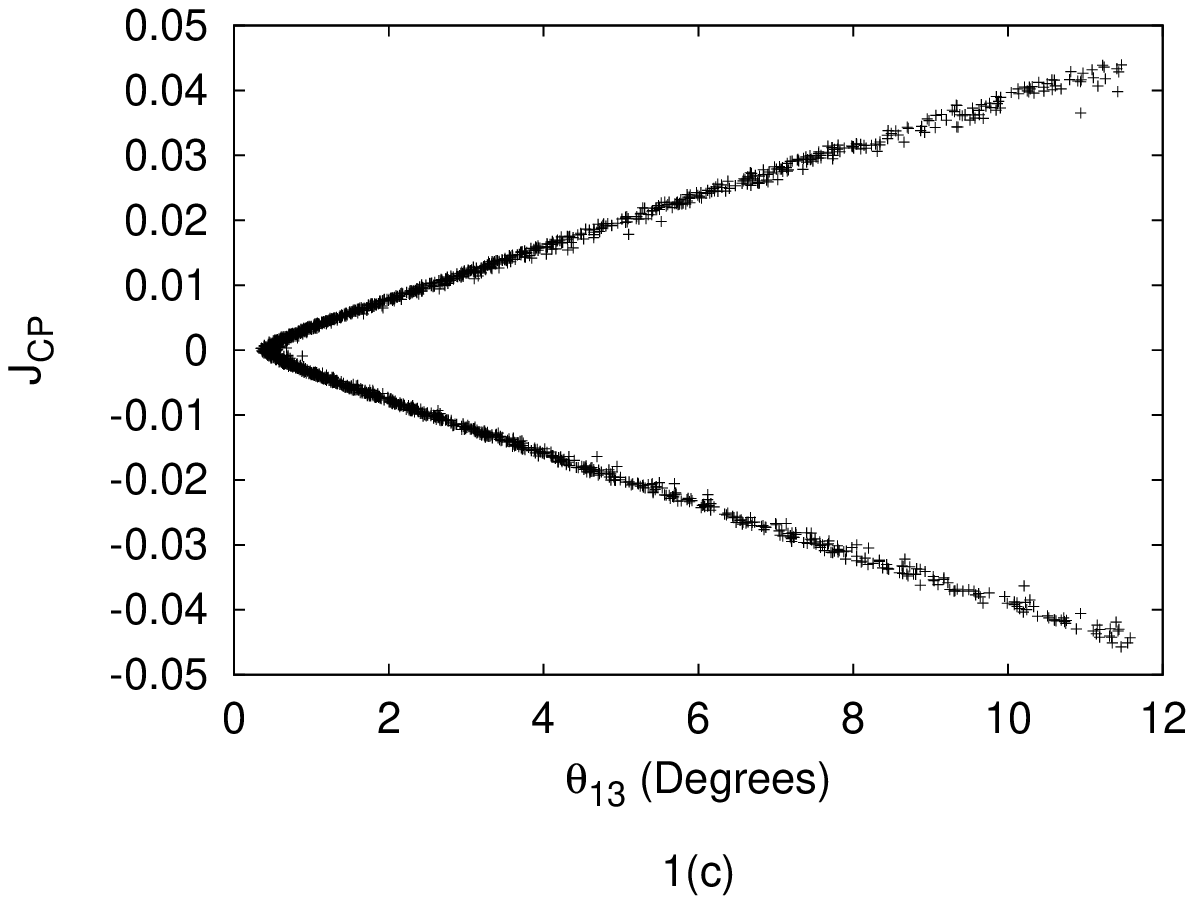,height=5.0cm,width=5.0cm}
\epsfig{file=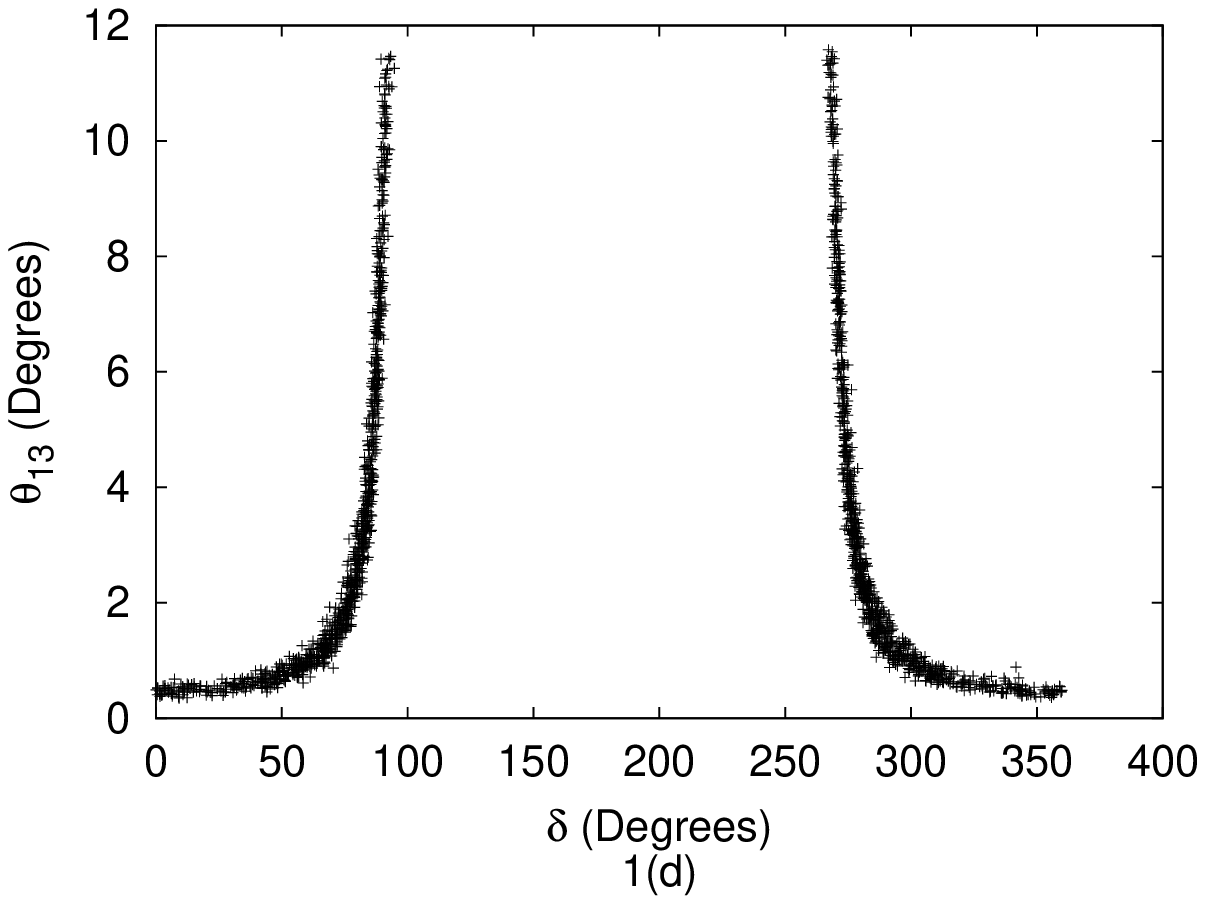,height=5.0cm,width=5.0cm}
\end{center}
\caption{Correlation plots for class 2A.}
\end{figure}

\begin{figure}
\begin{center}
\epsfig{file=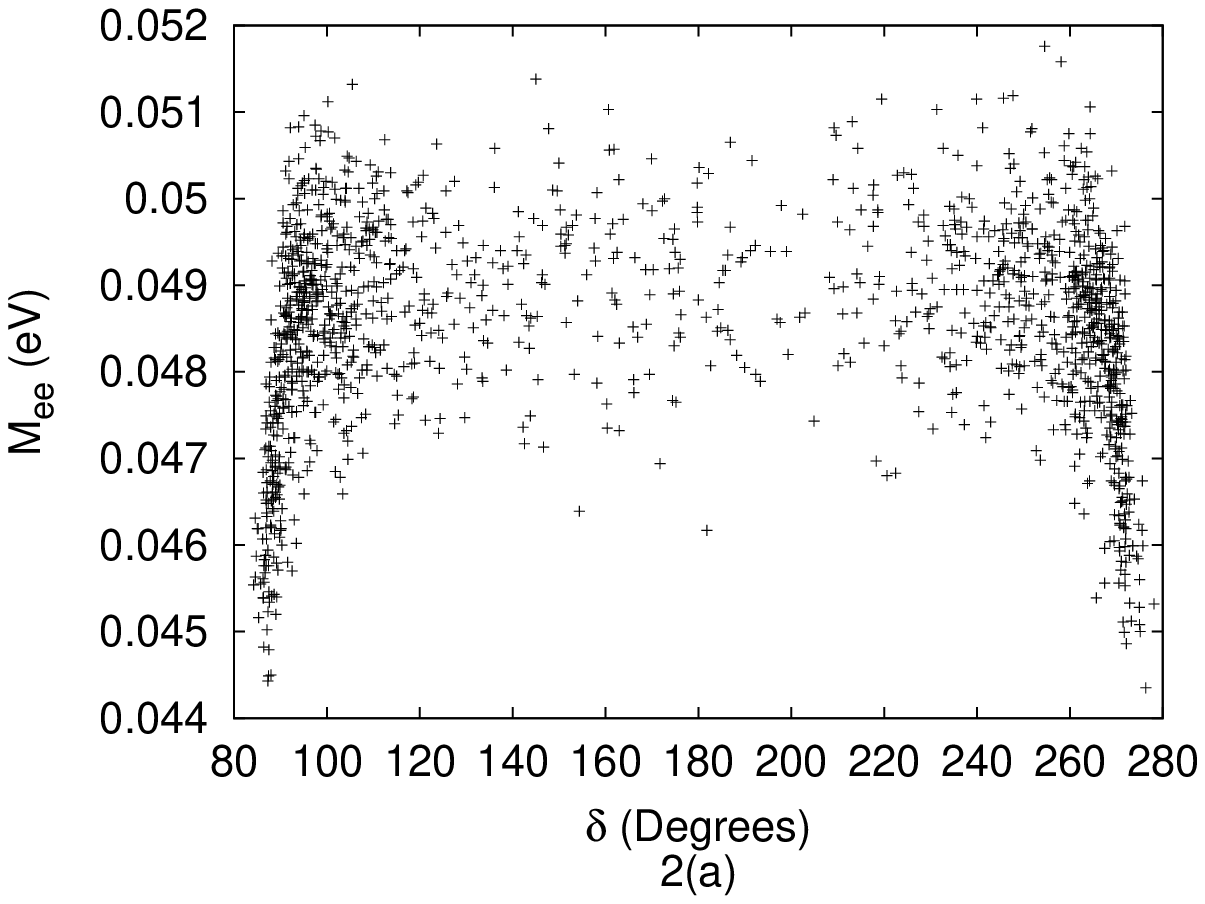,height=5.0cm,width=5.0cm}
\epsfig{file=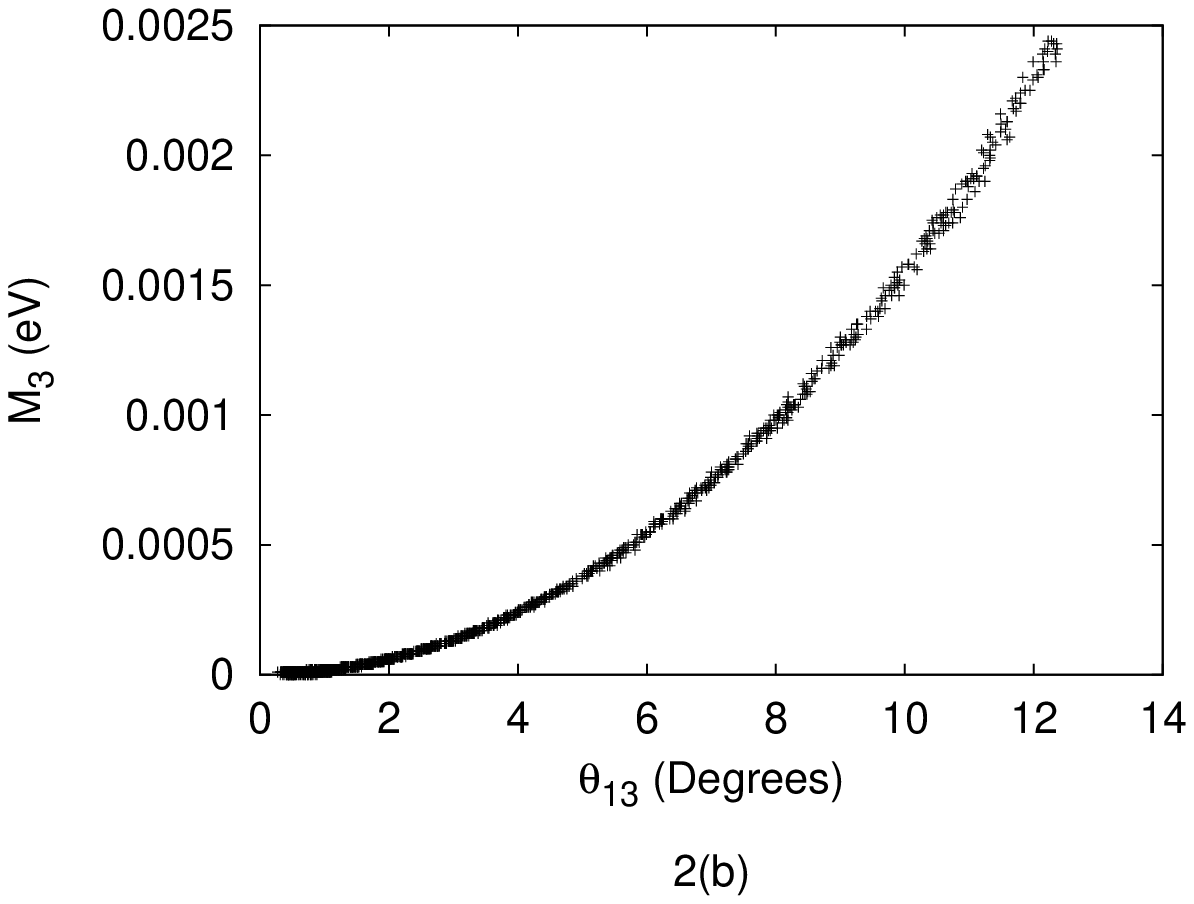,height=5.0cm,width=5.0cm} \caption{Correlation
plots for class 3A.} \epsfig{file=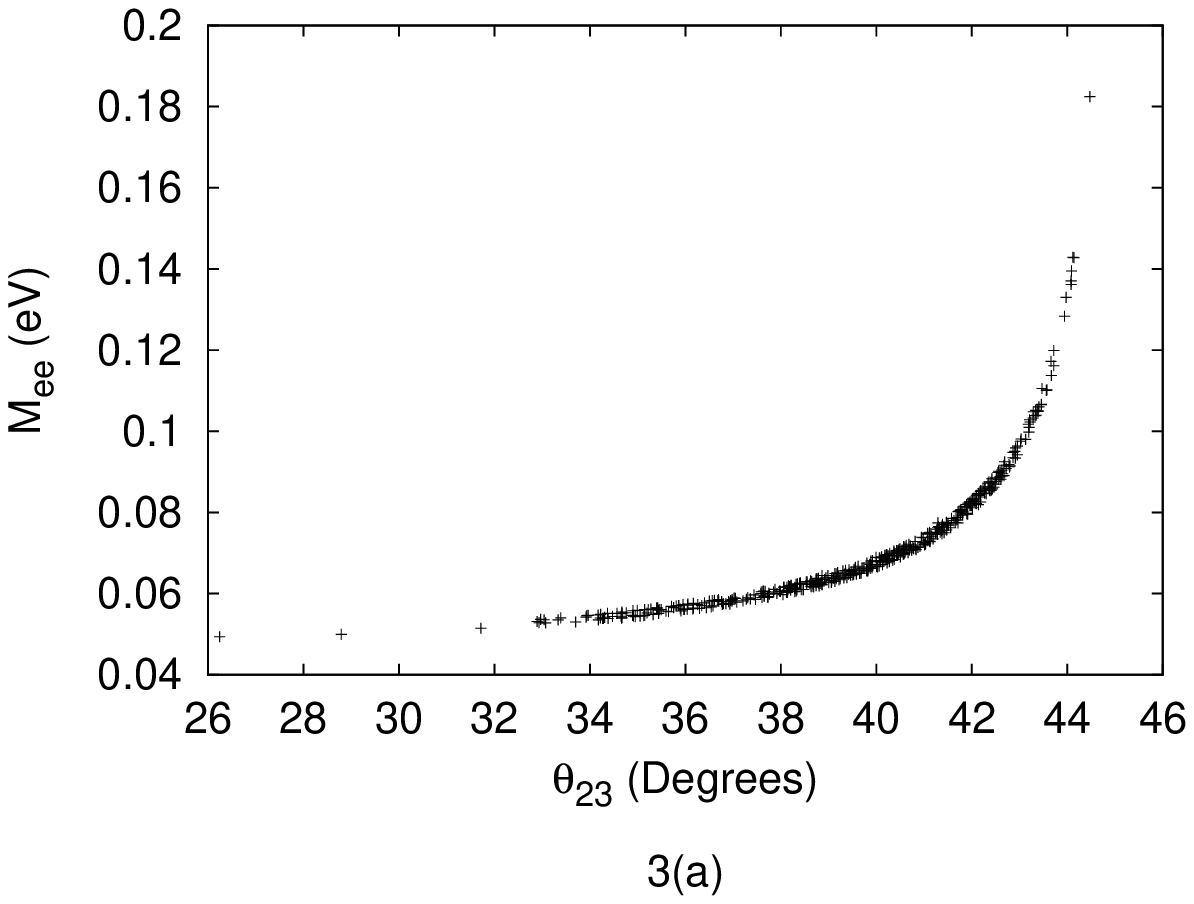,height=5.0cm,width=5.0cm}
\epsfig{file=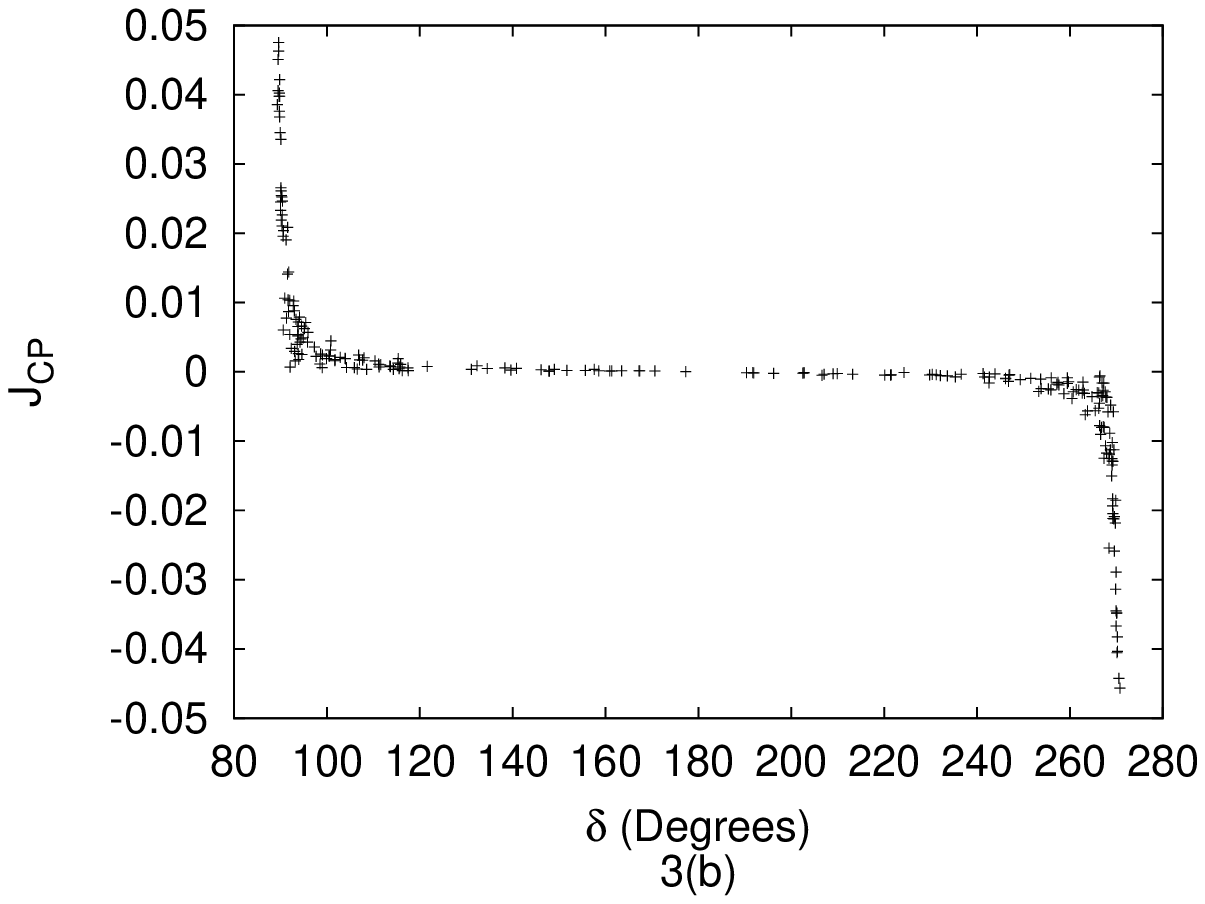,height=5.0cm,width=5.0cm}
\epsfig{file=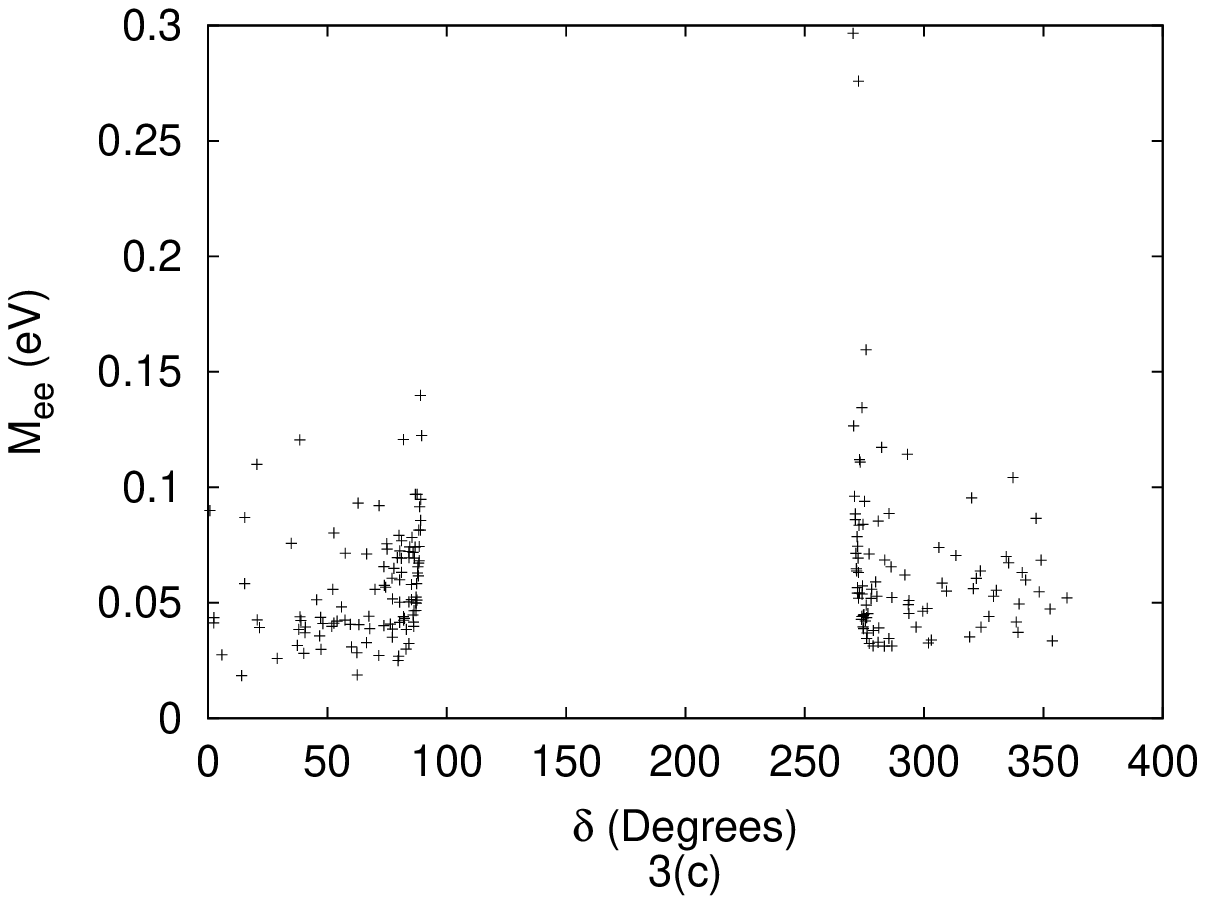,height=5.0cm,width=5.0cm} \caption{ Correlation
plots for class 2D (IH), 3F (IH) and 6C (NH) respectively.}
\end{center}
\end{figure}

\end{document}